\documentclass{llncs}
\pdfoutput=1
\usepackage{url}
\usepackage{amsmath}
\usepackage{xcolor}
\usepackage{proof}
\usepackage{argoclp}
\usepackage{tikz}
\usepackage{coqdoc}
\usepackage{isabelle,isabellesym}

\usepackage[utf8]{inputenc}

\newcommand{\xml}{XML}

\begin{document}
\title{
A Vernacular for Coherent Logic\thanks{The first, second and the fourth author 
were partly supported by the Serbian-French Technology Co-Operation grant 
EGIDE/"Pavle Savi\'c" 680-00-132/2012-09/12. The first and the fourth author 
are partly supported by the grant ON174021 of the Ministry of Science of Serbia. The final publication is available at \url{http://link.springer.com}.}}
\author{Sana Stojanovi\'c\inst{1} 
\and Julien Narboux\inst{2}
\and Marc Bezem\inst{3}
\and Predrag Jani\v{c}i\'c\inst{1}}
\institute{Faculty of Mathematics, University of Belgrade, 
%Studentski trg 16, 11000 Belgrade, 
Serbia 
\and
ICube, UMR 7357 CNRS, University of Strasbourg, 
% P\^ole API, Bd S\'ebastien Brant, BP 10413, 67412 Illkirch, 
France\\
\and
Institute for Informatics, University of Bergen,
% Postboks 7800, N-5020 Bergen, 
Norway \\ 
\email{sana@matf.bg.ac.rs,narboux@unistra.fr,\\marc.bezem@ii.uib.no,janicic@matf.bg.ac.rs}
}

\maketitle

\begin{abstract}
We propose a simple, yet expressive proof representation from which 
proofs for different proof assistants can easily be generated. The 
representation uses only a few inference rules and is based on a 
fragment of first-order logic called coherent logic. Coherent logic
has been recognized by a number of researchers as a suitable logic 
for many everyday mathematical developments. The proposed proof
representation is accompanied by a corresponding \xml{} format and 
by a suite of XSL transformations for generating formal proofs 
for Isabelle/Isar and Coq, as well as proofs expressed in a natural 
language form (formatted in \LaTeX{} or in HTML). Also, our automated 
theorem prover for coherent logic exports proofs in the proposed \xml{} 
format. All tools are publicly available, along with a set of sample 
theorems.
\end{abstract}

% ***************************************************************************
\section{Introduction}
\label{sec:introduction}
% ***************************************************************************

Mathematics can be done on two different levels. One level is rather informal, 
based on informal explanations, intuition, diagrams, etc., and typical for 
everyday mathematical practice. Another level is formal mathematics with 
proofs rigorously constructed by rules of inference from axioms. A large 
portion of mathematical logic and interactive theorem proving is aimed at 
linking these two levels. However, there is still a big gap: mathematicians 
still don't feel comfortable doing mathematics formally and proof assistants
still don't provide enough support for dealing with large mathematical
theories, automating technical problems, translating from one formalism to 
another, etc. We consider the following issue: there are several 
very mature and popular interactive theorem provers (including Isabelle, 
Coq, Mizar, HOL-light, see \cite{SeventeenProvers} for an overview), 
but they still cannot easily share the same mathematical knowledge. 
This is a significant problem, because there are increasing efforts 
in building repositories of formalized mathematics, but --- still 
developed within specific proof assistants. Building a mechanism for 
translation between different proof assistants is non-trivial 
because of many deep specifics of each proof assistant (there are 
some recent promissing approaches for this task \cite{KaliszykK13}). 
Instead of developing a translation mechanism, we propose a proof 
representation and a corresponding \xml-based format.
The proposed proof representation is light-weight and it does not aim 
at covering full power of everyday mathematical proofs or full power 
of first order logic. Still, it can cover a significant portion of many 
interesting mathematical theories. 
The underlying logic of our representation is coherent logic, a fragment 
of first-order logic. Proofs in this format can be generated in an easy way 
by dedicated, coherent logic provers, but in principle, also by standard 
theorem provers. The proofs can be translated to a range of proof 
assistant formats, enabling sharing the same developments. 

We call our proof representation ``coherent logic vernacular''. {\em Vernacular}
is the everyday, ordinary language (in contrast to the official, literary 
language) of the people of some country or region. A similar term, 
\emph{mathematical vernacular} was used in 1980's by de Bruijn within his 
formalism proposed for trying to \emph{put a substantial part of the mathematical 
vernacular into the formal system} \cite{deBruijnVernacular}. Several 
authors later modified or extended de Bruijn's framework. Wiedijk follows
de Bruijn's motivation \cite{Wiedijk2000-vernacular}, but he also notices:

\begin{quote}
\emph{It turns out that in a significant number of systems (‘proof assistants’)
one encounters languages that look almost the same. Apparently there is a 
canonical style of presenting mathematics that people discover independently: 
something like a {\em natural} mathematical vernacular. Because this language 
apparently is something that people arrive at independently, we might call 
it {\em the} mathematical vernacular.}
\end{quote}

We find that this language is actually closely related to a proof language 
of coherent logic, which is a basis of our proof representation presented 
in this paper.

Our proof representation is developed also with {\em readable proofs} in mind. 
Readable proofs (e.g., textbook-like proofs), are  very important in mathematical 
practice. For mathematicians, the main goal is often, not only a trusted, but 
also a clear and intuitive proof. We believe that coherent logic is very well 
suited for automated theorem proving with a simple production of readable proofs.

% ***************************************************************************
\section{Background}
% ***************************************************************************

In this section, we give a brief overview of interactive theorem proving and
proof assistants, of coherent logic, which is the logical basis
for our proof representation, and of \xml, which is the technical 
basis for our proof format.

% ---------------------------------------------------------------------------
\subsection{Interactive Theorem Proving}
% ---------------------------------------------------------------------------

Interactive theorem proving systems (or {\em proof assistants}) support 
the construction of formal proofs by a human, and verify each proof step 
with respect to the given underlying logic. The proofs can be written either
in a \emph{declarative} or in a \emph{procedural} proof style.
In the procedural proof style, the proof is described by a sequence of commands
 which modify the incomplete proof tree. In the declarative proof style the formal
document includes the intermediate statements.
Both styles are avaible in HOL-Light, Isabelle~\cite{Wenzel99} and Coq proof assistants 
whereas only the declarative style is available in Mizar, 
see~\cite{Wiedijk2012} for a recent discussion.
 The procedural proof style is more popular in the Coq community.
 
Formal proofs are typically much longer than 
``traditional proofs''.\footnote{The ratio between the length of formal proof 
\emph{script} and the length of the informal proof is often called the 
\emph{de Bruijn factor} \cite{Barendregt2005}. 
It varies for different parts of mathematics and for different systems,
and is currently often around 4. The de Bruijn factor can be below 1 if a lot
of automation can be used. It can also be well over 10 when the informal proof
is rather sketchy.} Progress in the field can be measured by proof scripts
becoming shorter and yet contain enough information for the system
to construct and verify the full (formal) proof.
``Traditional proofs'' can often hardly be called proofs, 
because of the many missing parts, informal arguments, etc. Using interactive 
theorem proving uncovered many flaws in many published mathematical proofs 
(including some seminal ones), published in books and journals.  

% I comment this because it is specific to Isar the comment above is more general
%The Intelligible semi-automated reasoning (Isar) \cite{Wenzel99} is 
%an alternative proof language interface layer for the Isabelle theorem 
%prover. The Isar language, with a level higher than traditional tactic scripts,
%bridges the gap between internal notions of interactive theorem provers 
%and proofs expressed in a human-friendly, readable way. The proofs in
%the Isar language are still machine verifiable.

% ---------------------------------------------------------------------------
\subsection{Coherent Logic}\label{subsec:cl}
% ---------------------------------------------------------------------------

Coherent logic (CL) was initially defined by Skolem and in recent years
it gained new attention \cite{CoherentLogic,SkolemMachines,HessenbergTheorem}.
It consists of formulae of the following form:

\begin{equation}
A_1(\vec{x}) \wedge \ldots \wedge A_n(\vec{x}) \Rightarrow
\exists \vec{y} ( B_1(\vec{x},\vec{y}) \vee \ldots \vee \; B_m(\vec{x},\vec{y}))
\label{eq:clf}
\end{equation}

\noindent
which are implicitly universally quantified, and where $0\leq n$, $0\leq m$,
$\vec{x}$ denotes a sequence of variables $x_1, x_2, \ldots, x_k$ ($0\leq k$),
$A_i$ (for $1 \leq i \leq n$) denotes an atomic formula (involving zero or more 
of the variables from $\vec{x}$), $\vec{y}$ denotes a sequence of variables
$y_1, y_2, \ldots, y_l$ ($0\leq l$), and $B_j$ (for $1 \leq j \leq m$)
denotes a conjunction of atomic formulae (involving zero or more of the variables 
from $\vec{x}$ and $\vec{y}$). 
For simplicity, we assume that there are no function symbols with arity greater 
than zero (so, we only consider symbols of constants as ground terms). 

The definition of CL does not involve negation. For a single atom $A$, $\neg A$ 
can be represented in the form $A \Rightarrow \bot$, where $\bot$ stands 
for the empty disjunction, but more general negation must
be expressed carefully in coherent logic. In order to reason with negation
in general, new predicate symbols are used to abbreviate subformulas.
Furthermore, for every predicate symbol $R$ (that appears in negated 
form), a new symbol $\overline{R}$ is introduced that stands for $\neg R$, 
and the following axioms are postulated (cf.\ \cite{PolonskyPhD}):
$\forall \vec{x} (R(\vec{x}) \wedge \overline{R}(\vec{x}) \Rightarrow \bot)$,
$\forall \vec{x} (R(\vec{x}) \vee \overline{R}(\vec{x}))$.

CL allows existential quantifications of the conclusion of a formula, 
so CL can be considered to be 
an extension of resolution logic. In contrast to the resolution-based proving, 
the conjecture being proved is kept unchanged and directly proved (refutation,
Skolemization and transformation to clausal form are not used). Hence,
proofs in CL are natural and intuitive and reasoning is constructive.
Readable proofs (in the style of forward reasoning and a variant of 
natural deduction) can easily be obtained \cite{CoherentLogic}.

A number of theories and theorems can be formulated directly and simply in CL.
In CL, constructive provability is the same as classical provability.
It can be proved that any first-order formula can be translated into a set
of CL formulas (in a different signature) preserving satisfiability 
\cite{PolonskyPhD} (however, this translation does not always 
preserve constructive provability).

Coherent logic is semi-decidable and there are several implemented semi-decision
procedures for it \cite{CoherentLogic}. ArgoCLP \cite{argoclp} is a generic 
theorem prover for coherent logic, based on a simple proof procedure with 
forward chaining and with iterative deepening. ArgoCLP  can read
problems given in TPTP form\footnote{\url{http://www.cs.miami.edu/~tptp/}} 
\cite{SS98} and can export proofs in the \xml{} format that we describe in 
this paper. These proofs are then translated into target languages, for
instance, the Isar language or natural language thanks to appropriate 
XSLT style-sheets.

% ---------------------------------------------------------------------------
\subsection{\xml}
% ---------------------------------------------------------------------------

{\em Extensible Markup Language} (\xml)\footnote{\url{http://www.w3.org/XML/}} 
is a simple, flexible text format, inspired by SGML (ISO 8879), 
for data structuring using tags and for interchanging information 
between different computing systems. \xml{} is primarily a ``metalanguage''---a 
language for describing other customized markup languages. So, it is not a 
fixed format like the markup language HTML---in \xml{} the tags 
indicate the semantic structure of the document, rather than only its layout. 
\xml{} is a project of the World Wide Web Consortium (W3C) and is a public 
format. Almost all browsers that are currently in use support \xml{} natively.

There are several schema languages for formaly specifying the structure 
and content of \xml{} documents of one class. Some of the main schema
languages are DTD ({\em Data Type Definition}), XML Schema, Relax, etc. 
\cite{LeeC00}. Specifications in the form of schema languages enable 
automatic verification (``validation'') of whether a specific document 
meets the given syntactical restrictions. 

{\em Extensible style-sheet language transformation} (XSLT)\footnote{\url{http://www.w3.org/Style/XSL/}}
is a document processing language that is used to transform the input
\xml{} documents to output files. An XSLT style-sheet 
declares a set of rules (templates) for an XSLT processor to 
use when interpreting the contents of an input \xml{} document. 
These rules tell to the {\scshape xslt} processor how that data should 
be presented: as an \xml{} document, as an {\scshape html} document, 
as plain text, or in some other form.

% ***************************************************************************
\section{Proof Representation}
\label{sec:proofrepresentation}
% ***************************************************************************

The proposed proof representation is very usable and expressive, yet very simple.
It uses only a few inference rules, a variant of the rules given in~\cite{HessenbergTheorem}.
Given a set of coherent axioms $AX$ and a coherent conjecture 
$A_1(\vec{x}) \wedge \ldots \wedge A_n(\vec{x}) \Rightarrow
\exists \vec{y} ( B_1(\vec{x},\vec{y}) \vee \ldots \vee \; B_m(\vec{x},\vec{y}))$,
the goal is to prove, using the rules given below, the following (where 
$\vec{a}$ denote a vector of new symbols of constants):
$$AX, A_1(\vec{a}) \wedge \ldots \wedge A_n(\vec{a}) \vdash 
\exists \vec{y} ( B_1(\vec{a},\vec{y}) \vee \ldots \vee \; B_m(\vec{a},\vec{y}))$$
The rules are applied in a forward manner, so they can be read from bottom to top.
In the rules below we assume:
\begin{itemize}
% \item $\Phi$ is of the form $\exists \vec{y} ( B_1(\vec{a},\vec{y}) \vee \ldots \vee \; B_m(\vec{a},\vec{y}))$;
  \item $ax\in AX$ is a formula of the form (\ref{eq:clf}) (page \pageref{subsec:cl});
 \item $\vec{a}$, $\vec{b}$, $\vec{c}$ denote vectors of constants (possibly of length zero);
 \item in the rule $\mathit{mp}$, $\vec{b}$ are fresh constants;
 \item $\vec{x}$ and $\vec{y}$ denote vectors of variables (possibly of length zero);
 \item $A_i(\vec{x})$ ($B_i(\vec{x},\vec{y})$) have no free variables other than from $\vec{x}$ (and $\vec{y}$);
 \item $A_i(\vec{a})$ are ground atomic formulae;
% \item $B_i(\vec{x},\vec{y})$ have no free variables other than from $\vec{x}$ and $\vec{y}$
 \item $B_i(\vec{a},\vec{b})$ and $B_i(\vec{c})$ are ground conjunctions of atomic formulae;
 \item $\underline{\Phi}$ denotes the list of conjuncts in $\Phi$.
 %, where $\vec{y}$ denotes a sequence $y_1, y_2, \ldots, y_k$ ($0\leq k$)
 \end{itemize}

$$\infer[\mathit{mp \; (modus \; ponens)}]
{\Gamma, ax, \underline{A_1(\vec{a}) \wedge \ldots \wedge A_n(\vec{a})} \vdash P}
{\Gamma, ax, \underline{A_1(\vec{a}) \wedge \ldots \wedge A_n(\vec{a})}, B_1(\vec{a},\vec{b}) \vee \ldots \vee B_m(\vec{a},\vec{b}) \vdash P}
$$

$$\infer[\mathit{cs \; (case \; split)}]
{\Gamma,B_1(\vec{c}) \vee \ldots \vee B_n(\vec{c}) \vdash P}
{\Gamma,\underline{B_1(\vec{c})} \vdash P   \,\,\,\,\,\,\,    \ldots \,\,\,\,\,\,\,  \Gamma, \underline{B_n(\vec{c})} \vdash P}
$$

$$\infer[\mathit{as \; (assumption)}]
{\Gamma, \underline{B_i(\vec{a},\vec{b})} \vdash \exists \vec{y} ( B_1(\vec{a},\vec{y}) \vee \ldots \vee \; B_m(\vec{a},\vec{y}))}
{}
$$

$$\infer[\mathit{efq \; (ex \; falso \; quodlibet)}]
{\Gamma,\bot \vdash P}
{}
$$
None of these rules change the goal $P$, which helps generating readable proofs as the goal can be kept implicit. 
Note that the rule $\mathit{mp}$ actually combines universal instantiation, %$\wedge$-introduction, 
conjunction introduction, modus ponens, and elimination of 
(zero or more) existential quantifiers. This seems a reasonable granularity for an inference step,
albeit probably the maximum for keeping proofs readable. 
Compared to~\cite{piotr87} which defines the notion of obvious inference rule by putting constraints on 
an automated prover, our position is: the obvious inferences are the ones defined by the inference rules above.
Compared to the rules given in~\cite{HessenbergTheorem}, we choose to separate the $\mathit{case \; split}$ rule 
(disjunction elimination) and the $\mathit{ex \; falso \; quodlibet}$ rule from the single combined rule in~\cite{HessenbergTheorem}, in order to improve readability.
Case distinction (split) is an important way of structuring proofs that deserves to be made explicit.
Also, $\mathit{ex \; falso\; quodlibet}$ could be seen as a $\mathit{case \; split}$ with zero cases,
but this would be less readable.

Any coherent logic proof can be represented in the following simple way 
($mp$ is used zero or more time, $\mathit{cs}$ involves at least two other $\mathit{proof}$ objects): 
$$\mathit{proof} \; ::= \; \mathit{mp}^* \;(\mathit{cs} (\mathit{proof}^{\ge 2}) \; | \; \mathit{as} \; | \; \mathit{efq} )$$

% ***************************************************************************
\section{\xml{} Suite for CL Vernacular}
\label{sec:implementation}
% ***************************************************************************

The proof representation described in Section \ref{sec:proofrepresentation}
is used as a basis for our XML-based proof format. It is developed as an 
interchange format for automated and interactive theorem provers. Proofs 
(for Coq and Isabelle/Isar) that are produced from our XML documents are 
fairly readable. The XML documents themselves can be read by a human, but 
much better alternative is using translation to human readable proofs in 
natural language (formatted in \LaTeX, for instance). The proof representation 
is described by a DTD \verb|Vernacular.dtd|. As an illustration, we show 
some fragments:

{\footnotesize
\begin{verbatim}
...
<!--******** Theory **************-->
<!ELEMENT theory (theory_name, signature, axiom*) >
<!ELEMENT theory_name (#PCDATA)>
<!ELEMENT signature (type*, relation_symbol*, constant*) >
<!ELEMENT relation_symbol (type*)>
<!ATTLIST relation_symbol name CDATA #REQUIRED>
<!ELEMENT type (#PCDATA)>
<!ELEMENT axiom (cl_formula)>
<!ATTLIST axiom name CDATA #REQUIRED>
...
\end{verbatim}
}

The above fragment describes the notion of theory. (Definitions, formalized 
as pairs of coherent formulae, are used as axioms.) A file describing a theory 
could be shared among several files with theorems and proofs.

{\footnotesize
\begin{verbatim}
...
<!--******** Theorem **************-->
<!ELEMENT theorem (theorem_name, cl_formula, proof+)>
<!ELEMENT theorem_name (#PCDATA)>
<!ELEMENT conjecture (name, cl_formula)>

<!--******** Proof **************-->
<!ELEMENT proof (proof_step*, proof_closing, proof_name?)>
<!ELEMENT proof_name EMPTY>
<!ATTLIST proof_name name CDATA #REQUIRED>

<!--******** Proof steps **************-->
<!ELEMENT proof_step (indentation,modus_ponens)>
<!ELEMENT proof_closing (indentation, (case_split|efq|from), 
	  (goal_reached_contradiction|goal_reached_thesis))>
...
\end{verbatim}
}

The above fragment describes the notion of a theorem and a proof. 
As said in Section \ref{sec:proofrepresentation}, a proof consists
of a sequence of applications of the rule modus ponens
and closes with one of the remaining proof rules ($\mathit{case \; split}$, 
$\mathit{as}$, or $\mathit{efq}$).
Within the last three, there is the additional information on
whether the proof closes by $\bot$ (by detecting a contradiction) 
or by detecting one of the disjuncts from the goal. This information
is generated by the prover and can be used for better readability
of the proof but also for some potential proof transformations.
Within each proof step there is also the information on indentation.
This information, useful for better layout, tells the level of 
subproofs and as such can be, in principle, computed from the \xml{} 
representation. Still, for convenience and simplicity of the XSLT 
style-sheets, it is stored within the \xml{} representation. 
 
We implemented XSL transformations from \xml{} format to Isabelle/Isar
(\verb|VernacularISAR.xls|), Coq (\verb|VernacularCoqTactics.xls|), and 
to a natural language (English) in \LaTeX{} form and in HTML form 
(\verb|VernacularTex.xls| and \verb|VernacularHTML.xls|). 

The translation from \xml{} to the Isar language is straightforward 
and each of our proof steps is trivially translated into Isar 
constructs.\footnote{The system Isabelle has available a proof method 
{\tt coherent} based on a internal theorem prover for coherent logic. 
Our Isar proofs do not use this proof method.} 
Naturally, we use native negation of Isar (and Coq) instead of defined 
negation in coherent logic. 

The translation to Coq has been written in the same spirit as the Isar 
output despite the fact proofs using tactics are more popular in Coq than 
declarative proofs. We refer to the assumptions by their statement instead 
of their name (for example: \verb|by cases on (A = B \/ A <> B)|). Moreover,  
when we can, we avoid to refer to the assumptions at all. 
We did not use the declarative proof mode of Coq because of efficiency issues.
We use our own tactics to implement the inference rules of CL to improve readability. Internally, we use an Ltac tactic to get the name of an assumption.
The forward reasoning proof steps consist of applications of the 
\verb|assert| tactic of Coq. Equality is translated into Leibniz equality. 

The translation to \LaTeX{} and HTML includes an additional XSLT style-sheet
that optionally defines specific layout for specific relation 
symbols (so, for instance, $(A,B) \cong (C,D)$ can be the layout 
for \verb|cong(A,B,C,D)|).

The developed XSLT style-sheets are rather simple and short --- 
each is only around 500 lines long. This shows that transformations
for other target languages (other theorem provers, like Mizar and HOL
light, \LaTeX{} with other natural languages, MathML, OMDoc or TPTP) 
can easily be constructed, thus enabling wide access to a 
single source of mathematical contents.

Our automated theorem prover for coherent logic ArgoCLP exports 
proofs in the form of the \xml{} files that conforms to this DTD. 
ArgoCLP reads an input theory and the conjecture given in the TPTP 
form (assuming the coherent form of all formulae and that there
are no function symbols or arity greater than 0).
ArgoCLP has built-in support for equality (during the search
process, it uses an efficient union-find structure) and the use of 
equality axioms is implicit in generated proofs.
The generated \xml{} documents are simple and consist of three
parts: \verb|frontpage| (providing, for instance, the author of the 
theorem, the prover used for generating the proof, the date),  
\verb|theory| (providing the signature and the axioms) and,
organized in chapters, a list of conjectures or theorems with 
their proofs. This way, some contents (\verb|frontpage| and
\verb|theory|) can be shared by a number of \xml{} documents. 
On the other hand, this also enables simple construction of 
bigger collections of theorems. The following is one example 
of an \xml{} document generated by ArgoCLP:
{\footnotesize
\begin{verbatim}
<?xml version="1.0" encoding="UTF-8"?>
<!DOCTYPE main SYSTEM "Vernacular.dtd">
<?xml-stylesheet href="VernacularISAR.xsl" type="text/xsl"?>

<main>
<xi:include href="frontpage.xml" parse="xml" 
    xmlns:xi="http://www.w3.org/2003/XInclude"/>
<xi:include href="theory_thm_4_19.xml" parse="xml" 
    xmlns:xi="http://www.w3.org/2003/XInclude"/>

<chapter name="th_4_19">
<xi:include href="proof_thm_4_19.xml" parse="xml" 
    xmlns:xi="http://www.w3.org/2003/XInclude"/>
</chapter>
</main>
\end{verbatim} 
}
\begin{figure}
\input{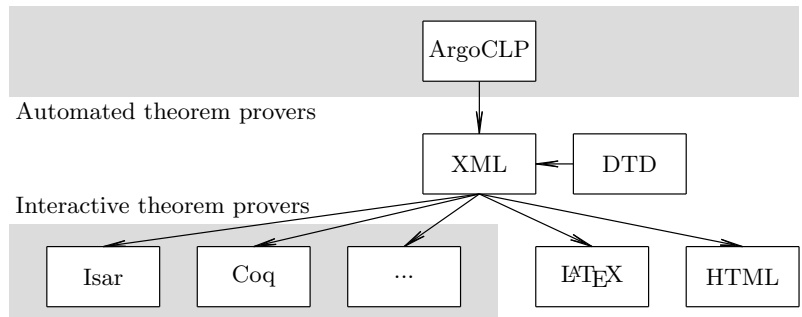}
\caption{Architecture of the presented framework}
\label{fig:architecture}
\end{figure}

The overall architecture of the framework is shown in Figure \ref{fig:architecture}.\footnote{The 
whole of our \xml{} suite, along with a collection of theorems is available online from 
\url{http://argo.matf.bg.ac.rs/downloads/software/clvernacular.zip}.}

% ***************************************************************************
\section{Examples}
\label{sec:examples}
% ***************************************************************************

Our \xml{} suite for coherent logic vernacular is used for a number of 
proofs generated by our prover ArgoCLP. In this section we 
discuss proofs of theorems from the book {\em Metamathematische Methoden in 
der Geometrie}, by Wolfram Schwabh\"auser, Wanda Szmielew, and Alfred Tarski 
\cite{tarski83}, one of the twenty-century mathematical classics. The theory 
is described in terms of first-order logic, it uses only one sort of primitive 
objects --- points, has only two primitive predicates ($cong$ or arity 4
and $bet$ of arity 3, intuitively for congruence and betweenness) and only 
eleven axioms. The majority of theorems from this book are in coherent logic 
or can be trivially transformed to belong to coherent logic. After needed 
transformations, the number of theorems in our development (238) is somewhat 
larger than in the book \cite{AutomatedTarski}.

Here we list a proof of one theorem (4.19) from Tarski's book. The theorem
was proved by ArgoCLP (using the list of relevant axioms and theorems 
produced by a resolution theorem prover), the proof was exported in the \xml{} format, and 
then transformed to a proof in natural language by appropriate XSL
transformation ($(A,B) \cong (C,D)$ is an infix notation for $cong(A,B,C,D)$
and it denotes that the pairs of points $(A,B)$ and $(C,D)$ are congruent,
$bet(A,B,C)$ denotes that the point $B$ is between the points $A$ and $C$,
$col(A,B,C)$ denotes that the points $A$, $B$ and $C$ are collinear).
%\hrulefill
\newcounter{proofstepnum}
\setcounter{proofstepnum}{0}

\begin{theorem}[th\_4\_19]
Assuming that $bet(A, B, C)$ and $AB \cong AD$ and $CB \cong CD$ it holds that $B = D$.
\end{theorem}

{\em Proof:}

\proofstep{0} {It holds that $bet(B, A, A)$ (using $th\_3\_1$).} 
\proofstep{0} {From the fact(s) $bet(A, B, C)$ it holds that $col(C, A, B)$ (using $ax\_4\_10\_3$).} 
\proofstep{0} {From the fact(s) $AB \cong AD$ it holds that $AD \cong AB$ (using $th\_2\_2$).} 
\proofstep{0} {It holds that $A = B$ or $A \neq B$.}  

\proofstep{3}  {Assume that: $A = B$.} 
\proofstep{6} {From the fact(s) $AD \cong AB$ and $A = B$ it holds that $AD \cong AA$.} 
\proofstep{6} {From the fact(s) $AD \cong AA$ it holds that $A = D$ (using $ax\_3$).} 
\proofstep{6} {From the fact(s) $A = B$ and $A = D$ it holds that $B = D$.} 

\proofstep{6} {The conclusion follows from the fact(s) $B = D$.}  

\proofstep{3}  {Assume that: $A \neq B$.} 
\proofstep{6} {It holds that $A = C$ or $A \neq C$.}  

\proofstep{9}  {Assume that: $A = C$.} 
\proofstep{12} {From the fact(s) $bet(A, B, C)$ and $A = C$ it holds that $bet(A, B, A)$.} 
\proofstep{12} {From the fact(s) $bet(A, B, A)$ and $bet(B, A, A)$ it holds that $A = B$ (using $th\_3\_4$).} 
\proofstep{12} {From the fact(s) $A \neq B$ and $A = B$ we get contradiction.}  
\proofstep{9}  {Assume that: $A \neq C$.} 
\proofstep{12} {From the fact(s) $A \neq C$ it holds that $C \neq A$.} 
\proofstep{12} {From the fact(s) $C \neq A$ and $col(C, A, B)$ and $CB \cong CD$ and $AB \cong AD$ it holds that $B = D$ (using $th\_4\_18$).} 

\proofstep{12} {The conclusion follows from the fact(s) $B = D$.} 
\proofstep{6} {The conclusion follows in all cases.}
\proofstep{0} {The conclusion follows in all cases.}

QED 
%\hrulefill

\newpage
Below is the same proof in Isabelle/Isar form: \\

\noindent\fbox{
\begin{minipage}{0.95\textwidth}
{%\footnotesize
\begin{isabellebody}%
\def\isabellecontext{theorem{\isacharunderscore}th{\isacharunderscore}{\isadigit{4}}{\isacharunderscore}{\isadigit{1}}{\isadigit{9}}}%

\isacommand{lemma}\isamarkupfalse%
\ th{\isacharunderscore}{\isadigit{4}}{\isacharunderscore}{\isadigit{1}}{\isadigit{9}}\ {\isacharcolon}\ \ \isakeyword{assumes}\ \ {\isachardoublequoteopen}bet\ A\ B\ C{\isachardoublequoteclose}\ \isakeyword{and}\ {\isachardoublequoteopen}cong\ A\ B\ A\ D{\isachardoublequoteclose}\ \isakeyword{and}\ {\isachardoublequoteopen}cong\ C\ B\ C\ D{\isachardoublequoteclose}\ \ \isakeyword{shows}\ {\isachardoublequoteopen}{\isacharparenleft}B\ {\isacharequal}\ D{\isacharparenright}{\isachardoublequoteclose}\ \isanewline
\isadelimproof
\isanewline
\endisadelimproof
\isatagproof
\isacommand{proof}\isamarkupfalse%
\ {\isacharminus}\ \isanewline
\isanewline
\isacommand{have}\isamarkupfalse%
\ {\isachardoublequoteopen}bet\ B\ A\ A{\isachardoublequoteclose}\ \ \isacommand{by}\isamarkupfalse%
\ {\isacharparenleft}rule\ th{\isacharunderscore}{\isadigit{3}}{\isacharunderscore}{\isadigit{1}}{\isacharparenright}\isanewline
\isacommand{from}\isamarkupfalse%
\ {\isacharbackquoteopen}bet\ A\ B\ C{\isacharbackquoteclose}\ \isacommand{have}\isamarkupfalse%
\ {\isachardoublequoteopen}col\ C\ A\ B{\isachardoublequoteclose}\ \isacommand{by}\isamarkupfalse%
\ {\isacharparenleft}rule\ ax{\isacharunderscore}{\isadigit{4}}{\isacharunderscore}{\isadigit{1}}{\isadigit{0}}{\isacharunderscore}{\isadigit{3}}{\isacharparenright}\isanewline
\isacommand{from}\isamarkupfalse%
\ {\isacharbackquoteopen}cong\ A\ B\ A\ D{\isacharbackquoteclose}\ \isacommand{have}\isamarkupfalse%
\ {\isachardoublequoteopen}cong\ A\ D\ A\ B{\isachardoublequoteclose}\ \isacommand{by}\isamarkupfalse%
\ {\isacharparenleft}rule\ th{\isacharunderscore}{\isadigit{2}}{\isacharunderscore}{\isadigit{2}}{\isacharparenright}\isanewline
\isacommand{have}\isamarkupfalse%
\ {\isachardoublequoteopen}A\ {\isacharequal}\ B\ {\isasymor}\ A\ {\isachartilde}{\isacharequal}\ B{\isachardoublequoteclose}\ \isacommand{by}\isamarkupfalse%
\ {\isacharparenleft}subst\ disj{\isacharunderscore}commute{\isacharcomma}\ rule\ excluded{\isacharunderscore}middle{\isacharparenright}\isanewline
\ \ \ \isacommand{show}\isamarkupfalse%
\ {\isacharquery}thesis\ \isanewline
\ \ \ \isacommand{proof}\isamarkupfalse%
{\isacharparenleft}cases\ {\isachardoublequoteopen}A\ {\isacharequal}\ B{\isachardoublequoteclose}{\isacharparenright}\isanewline
\ \ \ \ \ \isacommand{case}\isamarkupfalse%
\ True\isanewline
\ \ \ \ \ \ \isacommand{from}\isamarkupfalse%
\ {\isacharbackquoteopen}cong\ A\ D\ A\ B{\isacharbackquoteclose}\ \isakeyword{and}\ {\isacharbackquoteopen}A\ {\isacharequal}\ B{\isacharbackquoteclose}\ \isacommand{have}\isamarkupfalse%
\ {\isachardoublequoteopen}cong\ A\ D\ A\ A{\isachardoublequoteclose}\ \isacommand{by}\isamarkupfalse%
\ simp\isanewline
\ \ \ \ \ \ \isacommand{from}\isamarkupfalse%
\ {\isacharbackquoteopen}cong\ A\ D\ A\ A{\isacharbackquoteclose}\ \isacommand{have}\isamarkupfalse%
\ {\isachardoublequoteopen}A\ {\isacharequal}\ D{\isachardoublequoteclose}\ \isacommand{by}\isamarkupfalse%
\ {\isacharparenleft}rule\ ax{\isacharunderscore}{\isadigit{3}}{\isacharparenright}\isanewline
\ \ \ \ \ \ \isacommand{from}\isamarkupfalse%
\ {\isacharbackquoteopen}A\ {\isacharequal}\ B{\isacharbackquoteclose}\ \isakeyword{and}\ {\isacharbackquoteopen}A\ {\isacharequal}\ D{\isacharbackquoteclose}\ \isacommand{have}\isamarkupfalse%
\ {\isachardoublequoteopen}B\ {\isacharequal}\ D{\isachardoublequoteclose}\ \isacommand{by}\isamarkupfalse%
\ simp\isanewline
\ \ \ \ \ \ \isacommand{from}\isamarkupfalse%
\ {\isacharbackquoteopen}B\ {\isacharequal}\ D{\isacharbackquoteclose}\ \isacommand{show}\isamarkupfalse%
\ {\isacharquery}thesis\ \isacommand{by}\isamarkupfalse%
\ assumption\isanewline
\ \ \ \isacommand{next}\isamarkupfalse%
\isanewline
\ \ \ \ \ \isacommand{case}\isamarkupfalse%
\ False\isanewline
\ \ \ \ \ \ \isacommand{have}\isamarkupfalse%
\ {\isachardoublequoteopen}A\ {\isacharequal}\ C\ {\isasymor}\ A\ {\isachartilde}{\isacharequal}\ C{\isachardoublequoteclose}\ \isacommand{by}\isamarkupfalse%
\ {\isacharparenleft}subst\ disj{\isacharunderscore}commute{\isacharcomma}\ rule\ excluded{\isacharunderscore}middle{\isacharparenright}\isanewline
\ \ \ \ \ \ \ \ \ \isacommand{show}\isamarkupfalse%
\ {\isacharquery}thesis\isanewline
\ \ \ \ \ \ \ \ \ \isacommand{proof}\isamarkupfalse%
{\isacharparenleft}cases\ {\isachardoublequoteopen}A\ {\isacharequal}\ C{\isachardoublequoteclose}{\isacharparenright}\isanewline
\ \ \ \ \ \ \ \ \ \ \ \isacommand{case}\isamarkupfalse%
\ True\isanewline
\ \ \ \ \ \ \ \ \ \ \ \ \isacommand{from}\isamarkupfalse%
\ {\isacharbackquoteopen}bet\ A\ B\ C{\isacharbackquoteclose}\ \isakeyword{and}\ {\isacharbackquoteopen}A\ {\isacharequal}\ C{\isacharbackquoteclose}\ \isacommand{have}\isamarkupfalse%
\ {\isachardoublequoteopen}bet\ A\ B\ A{\isachardoublequoteclose}\ \isacommand{by}\isamarkupfalse%
\ simp\isanewline
\ \ \ \ \ \ \ \ \ \ \ \ \isacommand{from}\isamarkupfalse%
\ {\isacharbackquoteopen}bet\ A\ B\ A{\isacharbackquoteclose}\ \isakeyword{and}\ {\isacharbackquoteopen}bet\ B\ A\ A{\isacharbackquoteclose}\ \isacommand{have}\isamarkupfalse%
\ {\isachardoublequoteopen}A\ {\isacharequal}\ B{\isachardoublequoteclose}\ \isacommand{by}\isamarkupfalse%
\ {\isacharparenleft}rule\ th{\isacharunderscore}{\isadigit{3}}{\isacharunderscore}{\isadigit{4}}{\isacharparenright}\isanewline
\ \ \ \ \ \ \ \ \ \ \ \ \isacommand{from}\isamarkupfalse%
\ {\isacharbackquoteopen}A\ {\isachartilde}{\isacharequal}\ B{\isacharbackquoteclose}\ \isakeyword{and}\ {\isacharbackquoteopen}A\ {\isacharequal}\ B{\isacharbackquoteclose}\ \isacommand{have}\isamarkupfalse%
\ {\isachardoublequoteopen}False{\isachardoublequoteclose}\ \isacommand{by}\isamarkupfalse%
\ {\isacharparenleft}rule\ notE{\isacharparenright}\isanewline
\ \ \ \ \ \ \ \ \ \ \ \ \isacommand{from}\isamarkupfalse%
\ this\ \isacommand{show}\isamarkupfalse%
\ {\isacharquery}thesis\ \isacommand{by}\isamarkupfalse%
\ {\isacharparenleft}rule\ FalseE{\isacharparenright}\ \isanewline
\ \ \ \ \ \ \ \ \ \isacommand{next}\isamarkupfalse%
\isanewline
\ \ \ \ \ \ \ \ \ \ \ \isacommand{case}\isamarkupfalse%
\ False\isanewline
\ \ \ \ \ \ \ \ \ \ \ \ \isacommand{from}\isamarkupfalse%
\ {\isacharbackquoteopen}A\ {\isachartilde}{\isacharequal}\ C{\isacharbackquoteclose}\ \isacommand{have}\isamarkupfalse%
\ {\isachardoublequoteopen}C\ {\isachartilde}{\isacharequal}\ A{\isachardoublequoteclose}\ \isacommand{by}\isamarkupfalse%
\ {\isacharparenleft}rule\ not{\isacharunderscore}sym{\isacharparenright}\isanewline
\ \ \ \ \ \ \ \ \ \ \ \ \isacommand{from}\isamarkupfalse%
\ {\isacharbackquoteopen}C\ {\isachartilde}{\isacharequal}\ A{\isacharbackquoteclose}\ \isakeyword{and}\ {\isacharbackquoteopen}col\ C\ A\ B{\isacharbackquoteclose}\ \isakeyword{and}\ {\isacharbackquoteopen}cong\ C\ B\ C\ D{\isacharbackquoteclose}\ \isakeyword{and}\ {\isacharbackquoteopen}cong\ A\ B\ A\ D{\isacharbackquoteclose}\ \isacommand{have}\isamarkupfalse%
\ {\isachardoublequoteopen}B\ {\isacharequal}\ D{\isachardoublequoteclose}\ \isacommand{by}\isamarkupfalse%
\ {\isacharparenleft}rule\ th{\isacharunderscore}{\isadigit{4}}{\isacharunderscore}{\isadigit{1}}{\isadigit{8}}{\isacharparenright}\isanewline
\ \ \ \ \ \ \ \ \ \ \ \ \isacommand{from}\isamarkupfalse%
\ {\isacharbackquoteopen}B\ {\isacharequal}\ D{\isacharbackquoteclose}\ \isacommand{show}\isamarkupfalse%
\ {\isacharquery}thesis\ \isacommand{by}\isamarkupfalse%
\ assumption\isanewline
\ \ \ \ \ \ \ \ \ \isacommand{qed}\isamarkupfalse%
\isanewline
\ \ \ \isacommand{qed}\isamarkupfalse%
\isanewline
\isacommand{qed}\isamarkupfalse%
\endisatagproof
{\isafoldproof}%
\isadelimproof
\isanewline
\endisadelimproof
\isadelimtheory
\endisadelimtheory
\isatagtheory
\isacommand{end}\isamarkupfalse%
\endisatagtheory
{\isafoldtheory}%
\isadelimtheory
\isanewline
\endisadelimtheory
\end{isabellebody}%
}
\end{minipage}
}

\newpage
Below is the same proof in Coq form:

\noindent\fbox{
\begin{minipage}{0.95\textwidth}
\begin{coqdoccode}
\coqdockw{Theorem} \coqdef{theorem th 4 19.th 4 19}{th\_4\_19}{\coqdoclemma{th\_4\_19}} : \coqdockw{\ensuremath{\forall}} (\coqdocvar{A}:\coqref{theorem th 4 19.point}{\coqdocaxiom{point}}) (\coqdocvar{B}:\coqref{theorem th 4 19.point}{\coqdocaxiom{point}}) (\coqdocvar{C}:\coqref{theorem th 4 19.point}{\coqdocaxiom{point}}) (\coqdocvar{D}:\coqref{theorem th 4 19.point}{\coqdocaxiom{point}}), (\coqref{theorem th 4 19.bet}{\coqdocaxiom{bet}} \coqdocvariable{A} \coqdocvariable{B} \coqdocvariable{C} \coqexternalref{:type scope:x '/x5C' x}{http://coq.inria.fr/distrib/8.4pl3/stdlib/Coq.Init.Logic}{\coqdocnotation{\ensuremath{\land}}} \coqref{theorem th 4 19.cong}{\coqdocaxiom{cong}} \coqdocvariable{A} \coqdocvariable{B} \coqdocvariable{A} \coqdocvariable{D} \coqexternalref{:type scope:x '/x5C' x}{http://coq.inria.fr/distrib/8.4pl3/stdlib/Coq.Init.Logic}{\coqdocnotation{\ensuremath{\land}}} \coqref{theorem th 4 19.cong}{\coqdocaxiom{cong}} \coqdocvariable{C} \coqdocvariable{B} \coqdocvariable{C} \coqdocvariable{D}) 
\ensuremath{\rightarrow} \coqdocvariable{B} \coqexternalref{:type scope:x '=' x}{http://coq.inria.fr/distrib/8.4pl3/stdlib/Coq.Init.Logic}{\coqdocnotation{=}} \coqdocvariable{D}.\coqdoceol
\coqdocnoindent
\coqdockw{Proof}.\coqdoceol
\coqdocindent{0.50em}
\coqdoctac{intros}.\coqdoceol
\coqdocindent{0.50em}
\coqdoctac{assert} (\coqref{theorem th 4 19.bet}{\coqdocaxiom{bet}} \coqdocvar{B} \coqdocvar{A} \coqdocvar{A}) \coqdoctac{by} \coqdocvar{applying} (\coqref{theorem th 4 19.th 3 1}{\coqdocaxiom{th\_3\_1}} \coqdocvar{B} \coqdocvar{A} ) .\coqdoceol
\coqdocindent{0.50em}
\coqdoctac{assert} (\coqref{theorem th 4 19.col}{\coqdocaxiom{col}} \coqdocvar{C} \coqdocvar{A} \coqdocvar{B}) \coqdoctac{by} \coqdocvar{applying} (\coqref{theorem th 4 19.ax 4 10 3}{\coqdocaxiom{ax\_4\_10\_3}} \coqdocvar{A} \coqdocvar{B} \coqdocvar{C} ) .\coqdoceol
\coqdocindent{0.50em}
\coqdoctac{assert} (\coqref{theorem th 4 19.cong}{\coqdocaxiom{cong}} \coqdocvar{A} \coqdocvar{D} \coqdocvar{A} \coqdocvar{B}) \coqdoctac{by} \coqdocvar{applying} (\coqref{theorem th 4 19.th 2 2}{\coqdocaxiom{th\_2\_2}} \coqdocvar{A} \coqdocvar{B} \coqdocvar{A} \coqdocvar{D} ) .\coqdoceol
\coqdocindent{0.50em}
\coqdoctac{assert} (\coqdocvar{A} \coqexternalref{:type scope:x '=' x}{http://coq.inria.fr/distrib/8.4pl3/stdlib/Coq.Init.Logic}{\coqdocnotation{=}} \coqdocvar{B} \coqexternalref{:type scope:x 'x5C/' x}{http://coq.inria.fr/distrib/8.4pl3/stdlib/Coq.Init.Logic}{\coqdocnotation{\ensuremath{\lor}}} \coqdocvar{A} \coqexternalref{:type scope:x '<>' x}{http://coq.inria.fr/distrib/8.4pl3/stdlib/Coq.Init.Logic}{\coqdocnotation{\ensuremath{\not=}}} \coqdocvar{B}) \coqdoctac{by} \coqdocvar{applying} (\coqref{theorem th 4 19.ax g1}{\coqdocaxiom{ax\_g1}} \coqdocvar{A} \coqdocvar{B} ) .\coqdoceol
\coqdocindent{0.50em}
\coqdoctac{by} \coqdocvar{cases} \coqdocvar{on} (\coqdocvar{A} \coqexternalref{:type scope:x '=' x}{http://coq.inria.fr/distrib/8.4pl3/stdlib/Coq.Init.Logic}{\coqdocnotation{=}} \coqdocvar{B} \coqexternalref{:type scope:x 'x5C/' x}{http://coq.inria.fr/distrib/8.4pl3/stdlib/Coq.Init.Logic}{\coqdocnotation{\ensuremath{\lor}}} \coqdocvar{A} \coqexternalref{:type scope:x '<>' x}{http://coq.inria.fr/distrib/8.4pl3/stdlib/Coq.Init.Logic}{\coqdocnotation{\ensuremath{\not=}}} \coqdocvar{B}).\coqdoceol
\coqdocindent{0.50em}
- \{ \coqdoceol
\coqdocindent{1.50em}
\coqdoctac{assert} (\coqref{theorem th 4 19.cong}{\coqdocaxiom{cong}} \coqdocvar{A} \coqdocvar{D} \coqdocvar{A} \coqdocvar{A})  \coqdoctac{by} (\coqdocvar{substitution}).\coqdoceol
\coqdocindent{1.50em}
\coqdoctac{assert} (\coqdocvar{A} \coqexternalref{:type scope:x '=' x}{http://coq.inria.fr/distrib/8.4pl3/stdlib/Coq.Init.Logic}{\coqdocnotation{=}} \coqdocvar{D}) \coqdoctac{by} \coqdocvar{applying} (\coqref{theorem th 4 19.ax 3}{\coqdocaxiom{ax\_3}} \coqdocvar{A} \coqdocvar{D} \coqdocvar{A} ) .\coqdoceol
\coqdocindent{1.50em}
\coqdoctac{assert} (\coqdocvar{B} \coqexternalref{:type scope:x '=' x}{http://coq.inria.fr/distrib/8.4pl3/stdlib/Coq.Init.Logic}{\coqdocnotation{=}} \coqdocvar{D})  \coqdoctac{by} (\coqdocvar{substitution}).\coqdoceol
\coqdocindent{1.50em}
\coqdocvar{conclude}.\coqdoceol
\coqdocindent{1.50em}
\}\coqdoceol
\coqdocindent{0.50em}
- \{ \coqdoceol
\coqdocindent{1.50em}
\coqdoctac{assert} (\coqdocvar{A} \coqexternalref{:type scope:x '=' x}{http://coq.inria.fr/distrib/8.4pl3/stdlib/Coq.Init.Logic}{\coqdocnotation{=}} \coqdocvar{C} \coqexternalref{:type scope:x 'x5C/' x}{http://coq.inria.fr/distrib/8.4pl3/stdlib/Coq.Init.Logic}{\coqdocnotation{\ensuremath{\lor}}} \coqdocvar{A} \coqexternalref{:type scope:x '<>' x}{http://coq.inria.fr/distrib/8.4pl3/stdlib/Coq.Init.Logic}{\coqdocnotation{\ensuremath{\not=}}} \coqdocvar{C}) \coqdoctac{by} \coqdocvar{applying} (\coqref{theorem th 4 19.ax g1}{\coqdocaxiom{ax\_g1}} \coqdocvar{A} \coqdocvar{C} ) .\coqdoceol
\coqdocindent{1.50em}
\coqdoctac{by} \coqdocvar{cases} \coqdocvar{on} (\coqdocvar{A} \coqexternalref{:type scope:x '=' x}{http://coq.inria.fr/distrib/8.4pl3/stdlib/Coq.Init.Logic}{\coqdocnotation{=}} \coqdocvar{C} \coqexternalref{:type scope:x 'x5C/' x}{http://coq.inria.fr/distrib/8.4pl3/stdlib/Coq.Init.Logic}{\coqdocnotation{\ensuremath{\lor}}} \coqdocvar{A} \coqexternalref{:type scope:x '<>' x}{http://coq.inria.fr/distrib/8.4pl3/stdlib/Coq.Init.Logic}{\coqdocnotation{\ensuremath{\not=}}} \coqdocvar{C}).\coqdoceol
\coqdocindent{1.50em}
- \{ \coqdoceol
\coqdocindent{2.50em}
\coqdoctac{assert} (\coqref{theorem th 4 19.bet}{\coqdocaxiom{bet}} \coqdocvar{A} \coqdocvar{B} \coqdocvar{A})  \coqdoctac{by} (\coqdocvar{substitution}).\coqdoceol
\coqdocindent{2.50em}
\coqdoctac{assert} (\coqdocvar{A} \coqexternalref{:type scope:x '=' x}{http://coq.inria.fr/distrib/8.4pl3/stdlib/Coq.Init.Logic}{\coqdocnotation{=}} \coqdocvar{B}) \coqdoctac{by} \coqdocvar{applying} (\coqref{theorem th 4 19.th 3 4}{\coqdocaxiom{th\_3\_4}} \coqdocvar{A} \coqdocvar{B} \coqdocvar{A} ) .\coqdoceol
\coqdocindent{2.50em}
\coqdoctac{assert} (\coqexternalref{False}{http://coq.inria.fr/distrib/8.4pl3/stdlib/Coq.Init.Logic}{\coqdocinductive{False}})  \coqdoctac{by} (\coqdocvar{substitution}).\coqdoceol
\coqdocindent{2.50em}
\coqdocvar{contradict}.\coqdoceol
\coqdocindent{2.50em}
\}\coqdoceol
\coqdocindent{1.50em}
- \{ \coqdoceol
\coqdocindent{2.50em}
\coqdoctac{assert} (\coqdocvar{C} \coqexternalref{:type scope:x '<>' x}{http://coq.inria.fr/distrib/8.4pl3/stdlib/Coq.Init.Logic}{\coqdocnotation{\ensuremath{\not=}}} \coqdocvar{A})  \coqdoctac{by} (\coqdocvar{substitution}).\coqdoceol
\coqdocindent{2.50em}
\coqdoctac{assert} (\coqdocvar{B} \coqexternalref{:type scope:x '=' x}{http://coq.inria.fr/distrib/8.4pl3/stdlib/Coq.Init.Logic}{\coqdocnotation{=}} \coqdocvar{D}) \coqdoctac{by} \coqdocvar{applying} (\coqref{theorem th 4 19.th 4 18}{\coqdocaxiom{th\_4\_18}} \coqdocvar{C} \coqdocvar{A} \coqdocvar{B} \coqdocvar{D} ) .\coqdoceol
\coqdocindent{2.50em}
\coqdocvar{conclude}.\coqdoceol
\coqdocindent{2.50em}
\}\coqdoceol
\coqdocindent{1.50em}
\}\coqdoceol
\coqdocnoindent
\coqdockw{Qed}.\coqdoceol
\coqdocemptyline
\end{coqdoccode}
\end{minipage}
}

\vspace{1em}
From the set of individual theorems (238), the prover ArgoCLP completely 
automatically proved 85 (36\%) of these theorems and generated proofs 
in the \xml{} format. We created a single \xml{} document that contains
all proved theorems and other theorems tagged as conjectures. 
The whole document matches the original book by Schwabh\"auser, 
Szmielew, and Tarski and can be explored in the \LaTeX{} (or PDF) form,
HTML or as Isabelle or Coq development.\footnote{
Translating the \xml{} document with 85 proofs by to Isabelle, Coq, HTML, 
\LaTeX{} (and then to PDF) takes altogether around 20s on a PC with AMD 
Opteron 6168. The resulting Isabelle document is verified in 30s, and 
the Coq document in 6s.}

% ***************************************************************************
\section{Related Work}
% ***************************************************************************

In \cite{Wiedijk2000-vernacular}, Wiedijk proposes a mathematical vernacular that
is in a sense the common denominator of the proof languages of Hyperproof, 
Mizar and Isabelle/Isar. We agree with his conclusion in the last
sentence of the quotation in the introduction, but we think that the 
three proof languages were \emph{not} discovered independently.
Natural deduction has been introduced by the Polish logicians
Łukasiewicz and Jaśkowski in the late 1920's, in reaction on the
formalisms of Frege, Russell and Hilbert. The term \emph{natural deduction} 
seems to have been used first by Gentzen, in German: 
\begin{quote}
    \emph{Ich wollte zunächst einmal einen Formalismus aufstellen, der dem 
    wirklichen Schließen möglichst nahe kommt. So ergab sich ein ``Kalkül des 
    natürlichen Schließens".}
    (First of all I wanted to set up a formalism that comes as close as possible to actual reasoning. 
    Thus arose a ``calculus of natural deduction".)—Gentzen, Untersuchungen über das logische 
    Schließen (Mathematische Zeitschrift 39, pp.176–210, 1935)
\end{quote}

The qualifier \emph{natural} was of course particularly well-chosen to 
express that the earlier formalisms were unnatural! As this was indeed the 
case, natural deduction quickly became the predominant logical system, helped 
by the seminal work by Gentzen on cut-elimination. (Ironically, 
this technical work in proof theory is best carried out with proofs 
represented in \emph{sequent calculus}, using natural deduction 
on the meta-level.)

It should thus not come as a surprise that the vernacular we propose also 
is based on natural deduction. One difference with Wiedijk's vernacular is 
that ours is based on coherent logic instead of full first-order logic. 
This choice is motivated in Section~\ref{subsec:cl} (easier semi-decision 
procedure and more readable proofs). Another difference is that 
Wiedijk allows proofs to be incomplete, whereas we stress complete proof
objects. This difference is strongly related to the fact that Wiedijk's 
vernacular is in the first place an input formalism for proof construction, whereas 
our vernacular is an output formalism for proof presentation and export of proofs 
to different proof assistants.
As far as we know, the mathematical vernacular proposed by Wiedijk's has 
not been implemented on its own, although Hyperproof, Mizar and Isabelle/Isar 
are developed using the same ideas.

A number of authors independently point to this or similar fragments of 
first-order logic as suitable for expressing significant portions of standard 
mathematics (or specifically geometry), for instance, Avigad et.al.~\cite{Avigad2009}
and Givant and Tarski et.al.~\cite{givant99,tarski83} in the context of a new 
axiomatic foundations of geometry. A recent paper by Ganesalingam and Gowers
\cite{GanesalingamG13} is also related to our work. Their goal is comparable 
to ours: full automation combined with human-style output. They propose 
inference rules which are very similar to our coherent logic based proof 
system. For example, their rule 
$\mathit{splitDisjunctiveHypothesis}$ corresponds to the rule $\mathit{case \; split}$, 
$\mathit{deleteDoneDisjunct}$ corresponds to $\mathit{as}$, 
$\mathit{removeTarget}$ corresponds to $\mathit{as}$ (with length of $\vec{{\bf y}}$ 
greater than 0), $\mathit{forwardsReasoning}$ corresponds to the rule $\mathit{mp}$. 
Yet, some rules they proposed are not part of our set of rules. The logic they use is 
full first-order, with 
a plan to include second-order features (this would also be perfectly 
possible for coherent logic, which is the first-order fragment of \emph{geometric} 
logic, which is in turn a fragment of higher-order logic, see \cite{Blass}).
Upon closer inspection, the paper by Ganesalingam and Gowers seems
to stay within the coherent fragment, and proofs by contraposition
and contradiction are delegated to future work. We find some support
for our approach in the observation by Ganesalingam and Gowers
that it will be hard to avoid that such reasoning patterns are applied
in ``inappropriate contexts''. On the other hand, the primary domain of
application of their approach  is metric space theory so far, with the 
ambition to attack problems in other
domains as well. It would be very interesting to test the two approaches
on the same problem sets. One difference is that \cite{GanesalingamG13}
insists on proofs being faithful to the thought processes, whereas we
would be happy if the prover finds a short and elegant proof even after
a not-so-elegant proof search. Another difference is that we are
interested in portability of proofs to other systems. To our knowledge,
the prover described in \cite{GanesalingamG13} is not publicly available.

Compared to OMDoc~\cite{Kohlhase06}, our proof format is much more specific 
(as we specify the inference rules we use) and has less features. It can be 
seen as a specific set of \texttt{methods} elements of the \texttt{derive} 
element of OMDoc. 

An alternative to using coherent logic provers would be using one of the 
more powerful automated theorem provers and exploiting existing and ongoing 
work on proof reconstruction and refactoring (see, for example, 
\cite{SmolkaB13,Blanchette13,KaliszykU13}).
This is certainly a viable option. However, reconstructing a proof from 
the log of a highly optimized prover is difficult. One problematic step 
is deskolemization, that is, proof reconstruction from a proof of the 
skolemized version of the problem. (The most efficient provers are based 
on resolution logic, and clausification including skolemizing is the first 
step in the solution procedure.) What can be said about this approach 
in its current stage is that more theorems can be proved, but their proofs 
can still be prohibitively complicated (or use additional axioms). It has 
been, however, proved beneficial to use powerful automated theorem provers 
as preprocessors, to provide hints for ArgoCLP. 

The literature contains many results about exchanging proofs between proof 
assistant using deep or shallow embeddings~\cite{obua06,KellerW10}. 
Boessplug, Cerbonneaux and Hermant propose to use the $\lambda\Pi$-calculus 
as a universal proof language which can express proof without losing 
their computational properties~\cite{MBoeQCarOHer12}.
To our knowledge, these works do not focus on the readability of proofs.

% ***************************************************************************
\section{Conclusions and Further Work}
\label{sec:conclusions}
% ***************************************************************************

Over the last years a lot of effort has been invested in combining the 
power of automated and interactive theorem proving: interactive theorem 
provers are now equipped with trusted support for SAT solving, SMT solving, 
resolution method, etc \cite{BlanchetteBN11,Sledgehammer2013}. 
These combinations open new frontiers for applications of theorem proving 
in software and hardware verification, but also in formalization of 
mathematics and for helping mathematicians in everyday practice. 
Exporting proofs in formats such as the presented one opens new possibilities
for exporting readable mathematical knowledge from automated 
theorem provers to interactive theorem provers. 
In the presented approach, the task of generating object-level 
proofs for proof assistants or proofs expressed in natural language is
removed from theorem provers (where it would be hard-coded) and,
thanks to the interchange \xml{} format, delegated to simple XSLT 
style-sheets, which are very flexible and additional XSLT style-sheets
(for additional target formats) can be developed without changing
the prover. Also, different automated theorem provers can benefit from 
this suite, as they don't have to deal with specifics of proof assistants.

The presented proof representation is not intended to serve as ``the 
mathematical vernacular''. However, it can cover a significant portion 
of many interesting mathematical theories while it is very simple. 

Often, communication between an interactive theorem prover and
an external automated theorem prover is supported by a verified,
trusted interface which enables direct calling to the prover. 
On the other hand, our work yields a common format which can
be generated by different automated theorem provers and from
which proofs for different interactive theorem provers can be
generated. The advantage of our approach relies on the fact 
that the proof which is exported is not just a certificate, 
it is meant to be human readable.

The current version of the presented XML suite does not support 
function symbols of arity greated than 0. For the future work, 
we are planning to add that support to the proof format and
to our ArgoCLP prover.

In the current version, for simplicity, the generated Isar and 
Coq proofs use tactics stronger than necessary. We will try to 
completely move to basic proofs steps while keeping simplicity 
of proofs. Beside planning to further improve existing XSLT 
style-sheets, we are also planning to implement support for 
additional target languages such as OMDoc.

\paragraph{Acknowledgements.} We are grateful to Filip Mari\'c
for his feedback and advices on earlier phases of this work.

%\bibliographystyle{plain}
%\bibliography{pj}

\begin{thebibliography}{10}

\bibitem{Avigad2009}
Jeremy Avigad, Edward Dean, and John Mumma.
\newblock {A Formal System for Euclid's Elements}.
\newblock {\em The Review of Symbolic Logic}, 2009.

\bibitem{Barendregt2005}
Henk Barendregt and Freek Wiedijk.
\newblock {The Challenge of Computer Mathematics}.
\newblock {\em Philosophical Transactions of the Royal Society},
  363(1835):2351--2375, 2005.

\bibitem{CoherentLogic}
Marc Bezem and Thierry Coquand.
\newblock {Automating Coherent Logic}.
\newblock In Geoff Sutcliffe and Andrei Voronkov, editors, {\em 12th
  International Conference on Logic for Programming, Artificial Intelligence,
  and Reasoning --- LPAR 2005}, volume 3835 of {\em Lecture Notes in Computer
  Science}. Springer-Verlag, 2005.

\bibitem{HessenbergTheorem}
Marc Bezem and Dimitri Hendriks.
\newblock {On the Mechanization of the Proof of Hessenberg's Theorem in
  Coherent Logic}.
\newblock {\em Journal of Automated Reasoning}, 40(1), 2008.

\bibitem{Blanchette13}
Jasmin~Christian Blanchette.
\newblock {Redirecting Proofs by Contradiction}.
\newblock In Jasmin~Christian Blanchette and Josef Urban, editors, {\em Third
  International Workshop on Proof Exchange for Theorem Proving, PxTP 2013, Lake
  Placid, NY, USA, June 9-10, 2013}, volume~14 of {\em EPiC Series}, pages
  11--26. EasyChair, 2013.

\bibitem{Sledgehammer2013}
Jasmin~Christian Blanchette, Sascha B{\"o}hme, and Lawrence~C. Paulson.
\newblock {Extending {S}ledgehammer with {SMT} Solvers}.
\newblock {\em Journal of Automated Reasoning}, 51(1):109--128, 2013.

\bibitem{BlanchetteBN11}
Jasmin~Christian Blanchette, Lukas Bulwahn, and Tobias Nipkow.
\newblock {Automatic Proof and Disproof in Isabelle/HOL}.
\newblock In Cesare Tinelli and Viorica Sofronie-Stokkermans, editors, {\em
  Frontiers of Combining Systems, 8th International Symposium, Proceedings},
  volume 6989 of {\em Lecture Notes in Computer Science}, pages 12--27.
  Springer, 2011.

\bibitem{Blass}
Andreas Blass.
\newblock {Topoi and Computation}.
\newblock {\em Bulletin of the EATCS}, 36:57--65, 1998.

\bibitem{MBoeQCarOHer12}
Mathieu Boespflug, Quentin Carbonneaux, and Olivier Hermant.
\newblock {The {$\lambda\Pi$}-calculus Modulo as a Universal Proof Language}.
\newblock In {\em Second Workshop on Proof Exchange for Theorem Proving
  (PxTP)}, volume 878 of {\em CEUR Workshop Proceedings}, pages 28--43.
  CEUR-WS.org, 2012.

\bibitem{deBruijnVernacular}
Nicolaas~Govert de~Bruijn.
\newblock {The Mathematical Vernacular, a Language for Mathematics with Typed
  Sets}.
\newblock In Dybjer et~al., editor, {\em Proceedings of the Workshop on
  Programming Languages}, 1987.

\bibitem{SkolemMachines}
John Fisher and Marc Bezem.
\newblock {Skolem Machines and Geometric Logic}.
\newblock In Cliff~B. Jones, Zhiming Liu, and Jim Woodcock, editors, {\em 4th
  International Colloquium on Theoretical Aspects of Computing --- ICTAC 2007},
  volume 4711 of {\em Lecture Notes in Computer Science}. Springer-Verlag,
  2007.

\bibitem{GanesalingamG13}
Mohan Ganesalingam and William~Timothy Gowers.
\newblock {A fully automatic problem solver with human-style output}.
\newblock {\em CoRR}, abs/1309.4501, 2013.

\bibitem{KaliszykK13}
Cezary Kaliszyk and Alexander Krauss.
\newblock {Scalable LCF-Style Proof Translation}.
\newblock In Sandrine Blazy, Christine Paulin-Mohring, and David Pichardie,
  editors, {\em Interactive Theorem Proving - 4th International Conference, ITP
  2013, Rennes, France, July 22-26, 2013. Proceedings}, volume 7998 of {\em
  Lecture Notes in Computer Science}, pages 51--66. Springer, 2013.

\bibitem{KaliszykU13}
Cezary Kaliszyk and Josef Urban.
\newblock {PRocH: Proof Reconstruction for HOL Light}.
\newblock In Maria~Paola Bonacina, editor, {\em Automated Deduction - CADE-24 -
  24th International Conference on Automated Deduction}, volume 7898 of {\em
  Lecture Notes in Computer Science}, pages 267--274. Springer, 2013.

\bibitem{KellerW10}
Chantal Keller and Benjamin Werner.
\newblock Importing {HOL Light} into {Coq}.
\newblock In {\em ITP}, pages 307--322, 2010.

\bibitem{Kohlhase06}
Michael Kohlhase.
\newblock {An OMDoc primer}.
\newblock In {\em OMDoc – An Open Markup Format for Mathematical Documents
  [version 1.2]}, volume 4180 of {\em Lecture Notes in Computer Science}, pages
  33--34. Springer Berlin Heidelberg, 2006.

\bibitem{LeeC00}
Dongwon Lee and Wesley~W. Chu.
\newblock Comparative analysis of six xml schema languages.
\newblock {\em SIGMOD Record}, 29(3):76--87, 2000.

\bibitem{obua06}
Steven Obua and Sebastian Skalberg.
\newblock {Importing HOL into Isabelle/HOL}.
\newblock In Ulrich Furbach and Natarajan Shankar, editors, {\em {Automated
  Reasoning}}, volume 4130 of {\em {Lecture Notes in Computer Science}}, page
  298–302. Springer Berlin Heidelberg, 2006.

\bibitem{PolonskyPhD}
Andrew Polonsky.
\newblock {\em Proofs, Types and Lambda Calculus}.
\newblock PhD thesis, University of Bergen, 2011.

\bibitem{piotr87}
Piotr Rudnicki.
\newblock {Obvious inferences}.
\newblock {\em Journal of Automated Reasoning}, 3(4):383–393, 1987.

\bibitem{tarski83}
Wolfram Schwabh\"auser, Wanda Szmielew, and Alfred Tarski.
\newblock {\em Metamathematische Methoden in der Geometrie}.
\newblock Springer-Verlag, Berlin, 1983.

\bibitem{SmolkaB13}
Steffen~Juilf Smolka and Jasmin~Christian Blanchette.
\newblock {Robust, Semi-Intelligible Isabelle Proofs from ATP Proofs}.
\newblock In Jasmin~Christian Blanchette and Josef Urban, editors, {\em Third
  International Workshop on Proof Exchange for Theorem Proving, PxTP 2013},
  volume~14 of {\em EPiC Series}, pages 117--132. EasyChair, 2013.

\bibitem{AutomatedTarski}
Sana Stojanovi\'c, Julien Narboux, and Predrag Jani\v{c}i\'c.
\newblock {Synergy Between Interactive and Automated Theorem Proving in
  Formalization of Mathematical Knowledge: A Case Study of Tarski's Geometry}.
\newblock {\em Submitted for publication}, 2014.

\bibitem{argoclp}
Sana Stojanovi\'c, Vesna Pavlovi\'c, and Predrag Jani{\v c}i{\' c}.
\newblock {A Coherent Logic Based Geometry Theorem Prover Capable of Producing
  Formal and Readable Proofs}.
\newblock In Pascal Schreck, Julien Narboux, and J\"urgen Richter-Gebert,
  editors, {\em Automated Deduction in Geometry}, volume 6877 of {\em Lecture
  Notes in Computer Science}. Springer, 2011.

\bibitem{SS98}
G.~Sutcliffe.
\newblock {The TPTP Problem Library and Associated Infrastructure: The FOF and
  CNF Parts, v3.5.0}.
\newblock {\em Journal of Automated Reasoning}, 43(4):337--362, 2009.

\bibitem{givant99}
Alfred Tarski and Steven Givant.
\newblock {Tarski's system of geometry}.
\newblock {\em The Bulletin of Symbolic Logic}, 5(2), June 1999.

\bibitem{Wenzel99}
Markus Wenzel.
\newblock {Isar - A Generic Interpretative Approach to Readable Formal Proof
  Documents}.
\newblock In Yves Bertot, Gilles Dowek, Andr{\'e} Hirschowitz, C.~Paulin, and
  Laurent Th{\'e}ry, editors, {\em Theorem Proving in Higher Order Logics
  (TPHOLs'99)}, volume 1690 of {\em Lecture Notes in Computer Science}, pages
  167--184. Springer, 1999.

\bibitem{Wiedijk2000-vernacular}
Freek Wiedijk.
\newblock Mathematical {V}ernacular.
\newblock Unpublished note. \url{http://www.cs.ru.nl/~freek/notes/mv.pdf},
  2000.

\bibitem{SeventeenProvers}
Freek Wiedijk, editor.
\newblock {\em The Seventeen Provers of the World}, volume 3600 of {\em Lecture
  Notes in Computer Science}. Springer, 2006.

\bibitem{Wiedijk2012}
Freek Wiedijk.
\newblock {A Synthesis of the Procedural and Declarative Styles of Interactive
  Theorem Proving}.
\newblock {\em Logical Methods in Computer Science}, 8(1), 2012.

\end{thebibliography}

\end{document}